# Multiscale mechanical study of the *Turritella terebra* and *Turritellinella tricarinata* seashells

Y. Liu[1], M. Lott[1], S.F. Seyyedizadeh[1], I. Corvaglia[1], G. Greco[2], V. F. Dal Poggetto[2], A.S. Gliozzi[1], R. Mussat Sartor[3], N. Nurra[3], C. Vitale-Brovarone[1], N. M. Pugno[2,4], F. Bosia[1], M. Tortello[1]

[1]Dipartimento di Scienza Applicata e Tecnologia (DISAT), Politecnico di Torino, 10129 Torino, Italy

[2]Laboratory for BioiBionspired, Bionic, Nano, Meta, Materials & Mechanics, Dipartimento di Ingegneria Civile, Ambientale e Meccanica, Università di Trento, 38123 Trento, Italy

[3]Dipartimento Scienze della Vita e Biologia dei Sistemi (DBIOS), Università degli Studi di Torino, 10123 Torino, Italy

[4]School of Engineering and Materials Science, Queen Mary University of London, London, United Kingdom

## Abstract

Marine shells are designed by nature to ensure mechanical protection from predators and shelter for mollusks living inside them. A large amount of work has been done to study the multiscale mechanical properties of their complex microstructure and to draw inspiration for the design of impact-resistant biomimetic materials. Less is known regarding the dynamic behavior related to their structure at multiple scales. Here, we present a combined experimental and numerical study of the shells of two different species of gastropod sea snail belonging to the Turritellidae family, featuring a peculiar helicoconic shape with hierarchical spiral elements. The proposed procedure involves the use of micro–Computed Tomography scans for the accurate determination of geometry, Atomic Force Microscopy and Nanoindentation to evaluate local mechanical properties, surface morphology and heterogeneity, as well as Resonant Ultrasound Spectroscopy coupled with Finite Element Analysis simulations to determine global modal behavior. Results indicate that the specific features of the considered shells, in particular their helicoconic and hierarchical structure, can also be linked to their vibration attenuation behavior. Moreover, the proposed investigation method can be extended to the study of other natural systems, to determine their structure-related dynamic properties, ultimately aiding the design of bioinspired metamaterials and of structures with advanced vibration control.

## 1. Introduction

It is well known that nature, over millions of years of evolution, has optimized the mechanical, optical, and thermal properties of biological systems. As far as mechanical properties are concerned, a rich literature can be found dealing with the optimized quasi-static properties of many natural systems (1–4). On the other hand, the dynamic mechanical properties of biological systems are, up to now, less explored, although a collection of several notable examples can be found in recent review papers (5,6).

Marine shells, and in particular their micro- and nano-structure, have long been studied as ideal material systems for mechanical protection (7), impact attenuation (8), and have also been a source of bioinspiration for impact-resistant materials (9). There are about 60,000/70,000 species of shell mollusks, the two largest classes being the Gastropoda and the Bivalvia. The molluscan shell is secreted by the mantle, or pallium, that is the epithelial layer covering the mollusks (10). The shell growth starts by the gland of a larval shell (the protoconch) and proceeds by the addition of new rings of conchiolin impregnated with calcium carbonate to the edge of the shell, and by addition of calcium carbonate over the whole of its inner surface and a small number of organic matrices (11). Calcium carbonate has three crystal polymorphs: calcite, aragonite, and vaterite, in addition to an amorphous form (12). Calcite is the most stable of the three polymorphs under ambient conditions, whereas aragonite is metastable and vaterite is the most unstable. Most mollusk shells consist of aragonite and/or calcite, whereas vaterite is rarely found. Usually, the secreted calcareous material is in the crystalline form of prisms of calcite arranged normally to the surface of the shell, whereas the inner part is made of layers of crystals of aragonite parallel to the shell surface. There are several types of microstructures, like the nacreous, the crossed lamellar, the foliated, and the prismatic layer microstructures (13–15). The calcium carbonate structural elements form at least 95% of the shell making them stiff and strong, while the remainder consists of biopolymeric material that helps to make them tough as well (16). Although the nacreous and prismatic layer structures are most extensively studied, the most abundant microstructure in mollusks is the cross laminar one. The thin lamellar prisms of about 100 nm thickness, called third-order lamellae, are aligned parallel and create the plate (second-order lamellae) that inclines at an angle of about 45° against the shell surface. This microstructure has a particularly high ceramic content (up to 99.9% in mass), yet it shows unusually good mechanical performance (17–19). Nacre features the highest tensile and compressive strength among the various architectures, while the crossed lamellar has the highest fracture toughness (7). Its functional features are intimately associated with the microstructure of the shell, that is usually formed by three layers: inner (close to the animal), middle, and outer. When the shell undergoes a quasi-static load caused, for example, by the bite of a predator, the outer layer experiences multiple cracks that are stopped at the interface with the middle layer. With increasing load, damage starts propagating also inside the middle layer, but bridging forces, created by the particular orientation of the internal lamellae, hinder crack propagation (20). Moreover, different microstructural responses have also been reported, depending on the velocity of the impacts, generating different mechanisms of attenuation and fracture formation showing in some cases a considerable enhancement of the dynamic fracture strength with respect to that of quasi-static loading (8).

While the microstructure in shells plays a fundamental role in impact attenuation, their overall shape also determines their modal characteristics and vibration damping capabilities. These can, at least partially, be related to the main function of the shell.

Numerous examples of efficient vibration damping structures exist in nature (6). Some well-known cases are the Mantis shrimp dactyl club (21), the woodpecker skull (22), and the turtle shell (23). A common feature of these examples, which is also found in seashells, is the multiscale (hierarchical) organization of the structure, from optimized microstructures combining stiff and compliant components, to vibration-deflecting or attenuating elements at intermediate scales (e.g. oriented fiber bundles, suture joints) (5,24–26), to tuned resonance properties at full scale, e.g., occurring far from typical working conditions. Ultrasonic techniques, in particular Resonant Ultrasound Spectroscopy (RUS) (27,28) have successfully helped to determine and investigate the macroscopic mechanical properties on several biological samples including wood (29,30), bones (31,32), or the human skull (33), also in combination with micro-computed tomography (µCT) scans and finite element analysis (34), while nano and microscale characterizations are also fundamental for addressing the same properties at smaller length scales (35–39).

In this paper, we focus on two shells belonging to the Turritellidae family, namely the *Turritella terebra* (Linnaeus, 1758) and the *Turritellinella tricarinata* (Brocchi, 1814). They are among the most common and abundant species on muddy and sandy seafloors, which are typical far from the coasts. Often, they are the dominant species, with hundreds of individuals per square meter. Both shells are characterized by a long tower-like shape which resembles a drill, as shown in Figure 1, hence the name. This long and narrow helicoconic shell shape may facilitate, through the foot, the access of the mollusk under the soft mud and anchoring to the marine substrate to prevent deep currents from carrying away the shell. Several structural features of *Turritella terebra* have been investigated in order to learn from nature, e.g., for the design of "adaptable structures", since this shell grows in an adaptive way and is able to change form through time to meet performance demands (40–42). Mathematical models of the shell have been proposed and finite element analysis has also been performed (42,43). The shells have three macroscopic layers, called the inner, middle, and outer layer, but it is not clear whether the calcium carbonate crystals are in the form of calcite or aragonite (42). Each is organized in the crossed lamellar architecture showing, from the largest to the smallest scale, the first, second and third order lamellae.

Here, we present a multiscale study of the mechanical properties of these two shells. By applying a combined numerical and experimental procedure at the micro and macro scale, we investigate the complex architecture of the shells and determine global mechanical properties studying their modal response and vibration properties. Their frequency response spectra are compared with simulated ones and discussed in terms of possible vibration attenuation capabilities, in the perspective of providing useful elements for the design bioinspired metamaterials with engineered wave propagation properties.

## 2. Materials and Methods

In the following, we will describe the techniques and procedures used for the investigation of the two shells. The µCT scan and the subsequent use of a specific software for the elaboration of the obtained images allowed to characterize the geometry of the shell that could be imported in a finite element software. The electron microscope was employed to directly determine the microstructure of the shells while the Atomic Force Microscopy (AFM) allowed to obtain the spatial distribution of the Young's modulus at the nanoscale along with the topography characterization of the polished surface of the samples. The obtained results were supported by further instrumented nanoindentation measurements. RUS experiments were performed to characterize the vibrational properties of the overall structure of the samples and numerical Finite Element Analysis (FEA) was used to retrieve the mechanical properties of the samples from the experimental results and to further investigate their vibration attenuation capabilities.

### 2.1. µCT-Scan and creation of a 3D mesh for FEA

CT scans of the shell were performed by using the three-dimensional (3D) X-ray microscope Skyscan 1272 by Bruker. The images allowed us to obtain a clear 3D view of the sample geometry. The outcome of this experiment was a triangulated STereoLithography (STL) file which was generated based on pixel density, as shown in Figure 1 b). STL files of the samples were prepared according to the procedure suggested by Bruker for the 3D scan. For the FEA, the STL files were then re-meshed to reduce the mesh density and imported into Meshmixer V3.5 software, where different operations such as Hole filling, Auto Repair, Mesh Smoothing, Mesh Solid and Remesh were used. This process was repeated several times to remove intersecting triangles, reduce the number of triangular elements and general errors in the file. The initial number of nodes was ranging between about 400 thousand and 1.5 million, depending on the case, and the density after re-meshing was 8-10% the initial one. The modified 3D STL file, reported in Figure 1 c), was then imported into a FEA software (COMSOL Multiphysics 6.0) for further simulation and analysis.

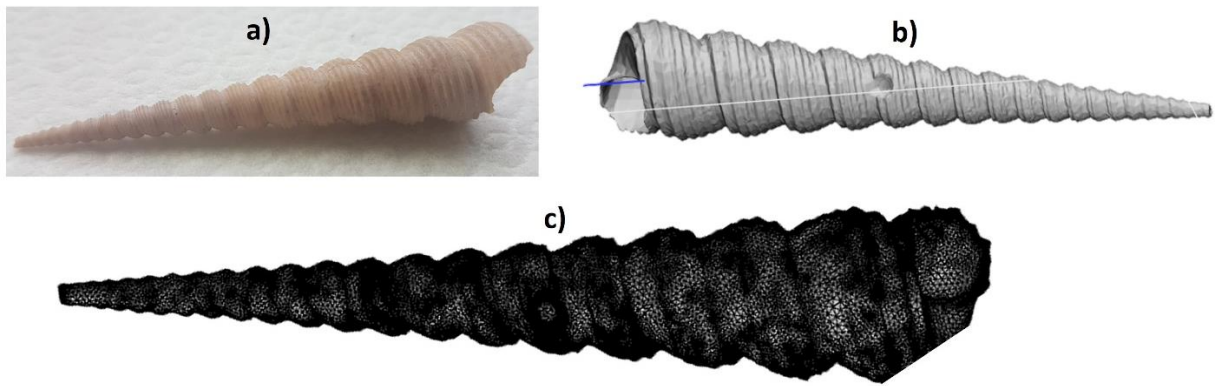

*Figure 1: a) Turritella terebra sample. b) 3D representation from STL file of Turritella terebra generated by micro-CT scan. c) Modified Turritella terebra STL file suitable for FEA.*

## 2.2. Scanning Electron Microscopy

SEM images of the sections of the two types of shell were acquired after mechanical fracturing of the samples. A Merlin field emission scanning electron microscope by Zeiss was used to obtain several images of the microstructure. Figure 2 shows some images of the *Turritella terebra* shell at different magnification levels. In panel a) it is possible to observe a layered structure formed by the first order lamellae, each with a thickness of a few microns (<10 µm). Further magnification, panel b), indicates that each layer is formed by lamellar elements with different orientations depending on the layer. Panels c) and d), show a magnification of the calcium carbonate lamellae forming the various layers. Each sub-element, corresponding to a third-order lamella, is approximately 100-150 nm thick.

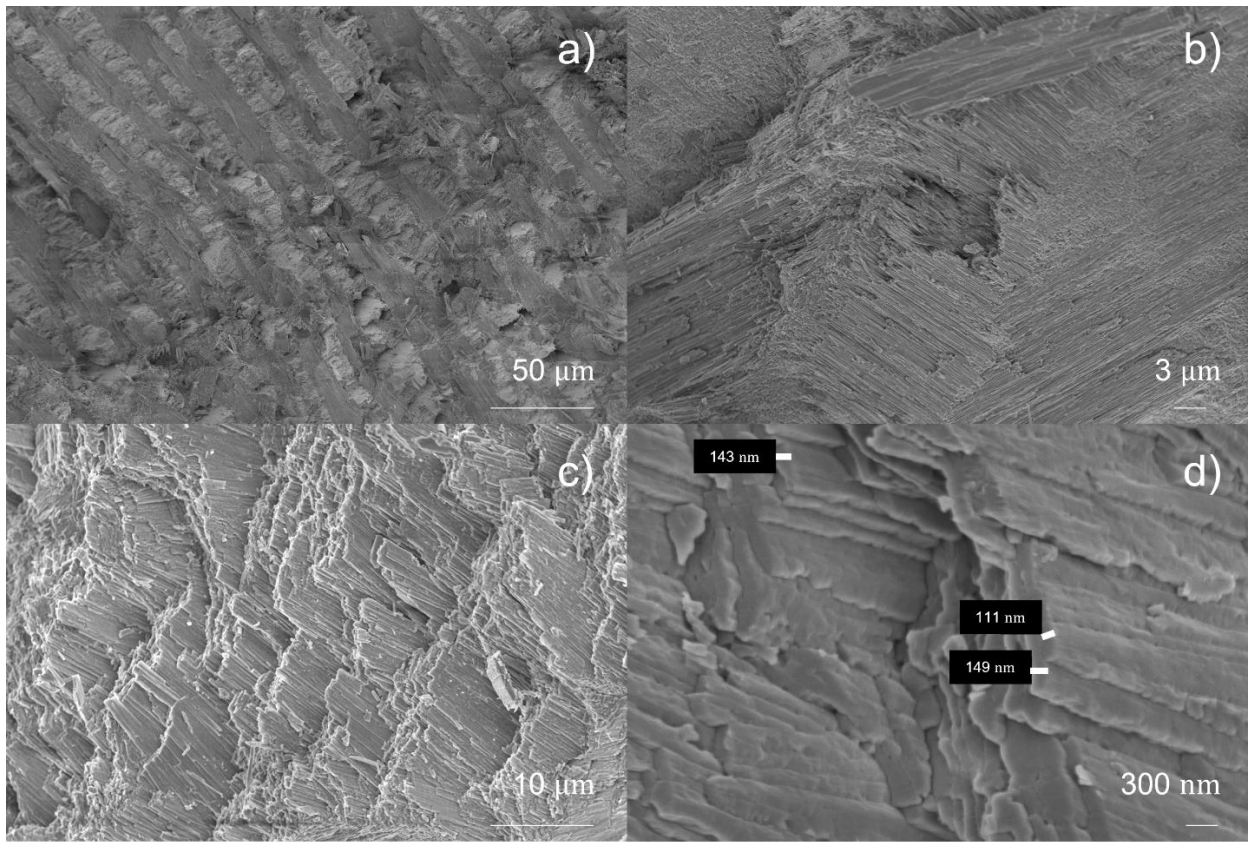

*Figure 2: SEM images of the* Turritella terebra *shell. a) first-order lamellae about* 10 μm *thick with alternating orientations of the second order lamellae. b) second-order lamellae stacked in groups forming different angles with respect to each other. c)-d) third-order lamellae with thickness ranging from about 100 nm to 150 nm.*

Similar images were acquired also for the *T. tricarinata* sample and are reported in the Supplemental material. The same cross laminar architecture was observed. Sometimes, as shown in Figure S1 c), some elements featuring a rod-like shape were also observed. However, in both cases, a clear, crossed lamellar, structured composition of the shell was revealed.

### 2.3. AFM

In order to measure the local mechanical properties of the *Turritella* shells, AFM experiments were conducted, allowing to obtain, by analyzing the force curves, the distribution of the elastic modulus on the sample surface. An Innova atomic force microscope by Bruker was used for the acquisition of the force vs distance curves. A hand-crafted ultra-high force probe (Bruker DNISP-HS) fabricated by precision grinding of a solid diamond with tip radius of 40 nm and tip half angle 42.3° was used. Sensitivity calibration of the AFM probe was done by using a sapphire sample and the reliability of the Young's modulus determination was verified on a fused silica test sample (Figure S2). The probe is specifically designed for nanoindentation and its stiffness, $k = 353$ N/m, is ideal for the expected range of Young's modulus of this sample (calcium carbonate).

*Turritella* samples were embedded in epoxy resin and polished with sandpaper up to 4000 grit, and subsequently with a 1 µm diamond paste to obtain a flat surface on which to acquire the force curves. Even if the roughness of the polished surface is of the order of 4 nm, the effect of the microstructure cannot be avoided as the sample itself is not a bulk material, but it is formed by calcium carbonate lamellae.

### 2.4. Instrumented Nanoindentation

To perform the tests on the shells, we used an established protocol (44). The shells were broken and fixed into an epoxy resin, which was left to dry for 24 h. The samples were then polished to achieve a roughness of about 40 nm. Three types of samples were produced: *Turritella terebra* (samples named TT0 and TT) and *Turritellinella tricarinata* (sample named TC1). We performed nanoindentation tests with the support of an iNano®Nanoindenter (Nanomechanics Inc.). The declared sensitivity of the machine is 3 nN for the load and 0.001 nm for the displacement. The shell was analysed by means of mapping method (Nanoblitz 3d, Nanomechanics Inc.). This gives the distribution of the mechanical properties of the sample on a specific squared area of interest (selected with the support of the coupled optical microscope). For each map, $40/50 \times 40/50$ points were investigated. A Berkovich tip, i.e. with a three-sided pyramidal shape, was mounted in the machine and used to perform the experiments (at controlled temperature and humidity, 20 °C and 30-40 % RH); the method used to compute the mechanical properties is the Oliver and Pharr (45), with indentation loads in the range 0.1-0.5 mN.

### 2.5. Ultrasonic measurements

RUS experiments were performed by applying chirp signals in the $1 - 45$ kHz frequency range to the TT sample, by using a 500-kHz Olympus piezoelectric transducer, as shown in Figure 3 a). Water gel was used as a coupling medium. The out-of-plane output vibration signal was recorded by means of a laser doppler vibrometer (Polytec) on 60 points along the longitudinal axis of the shell, previously carefully covered by reflective tape. The laser scan of the surface was automated by using a translational motor stage. Panel b) reports, as gray lines, the 60 spectra recorded during the measurements along with the average spectrum, indicated by the black line. The spectra differ mainly in the attenuation levels, in the level of noise, and relative spectral weight of the main peaks, but it is possible to recognize main common features that can be traced back to the entire structure. This justifies the study of the average spectrum of the sample. The average spectrum clearly shows several resonance modes of the structure that can be investigated further and suitably compared with FEA simulations.

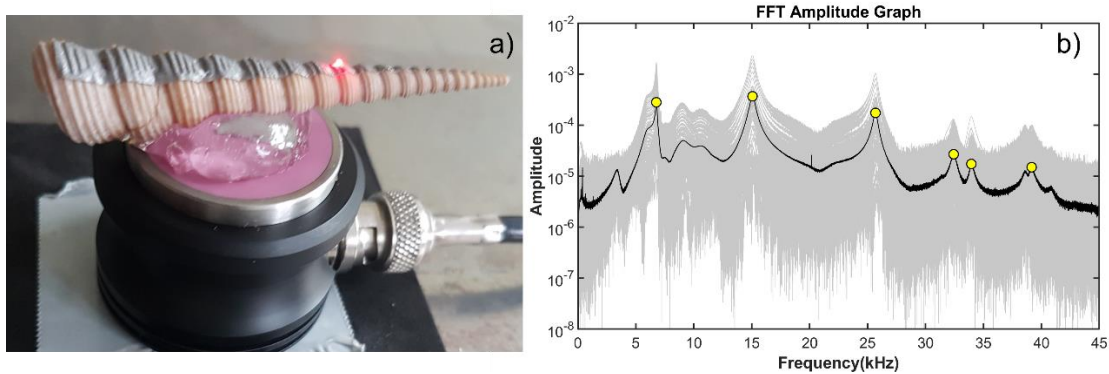

*Figure 3: a) TT sample mounted on the transducer during ultrasonic excitation experiments. b) Gray lines are the frequency spectra recorded along the longitudinal axis of the shell while the black line represents their average. Yellow symbols represent the resonance frequencies.*

## 3. Results

### 3.1. AFM results

Figure 4 a) shows a camera image of the polished *Turritella terebra* sample, called TT, and the microstructure is still visible on the right-hand side of the image, where the first order lamellae, like those shown in Figure 2 a), can be seen. A topography image of the polished surface, recorded on a flat area, is reported in panel b). The topography image mainly shows the edge between the outer (flat) and middle (more irregular) layer of the shell. The right-hand side of the surface shows some holes and voids in the microstructure, that were probably caused by the polishing procedure or by an intrinsic defect, but the remaining surface appears rather smooth. After recording the topography image, a matrix of $32 \times 32$ force vs distance curves was acquired in the area located by the blue square. The maximum force was set around 30 µN. Since the sample is hard and the adhesion force is very small compared to the maximum applied force and thus negligible, we fitted the approach curves by using the elastic Hertz model (46,47), as also done in (48).

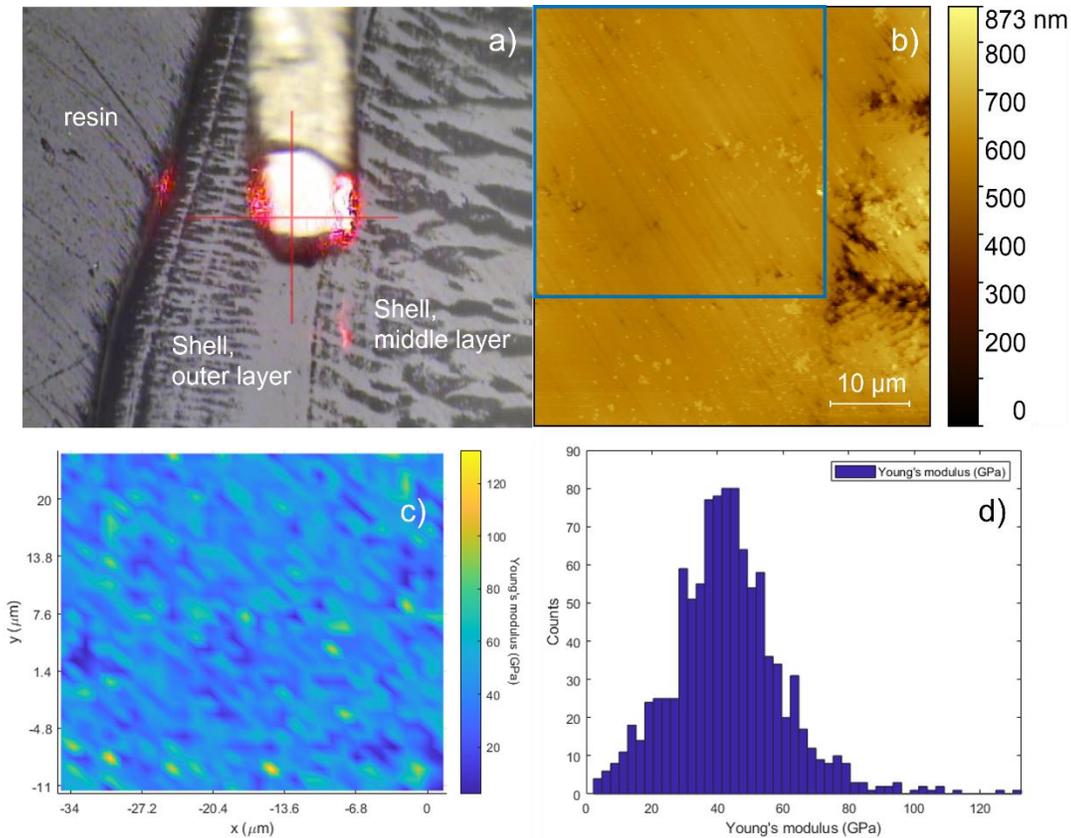

*Figure 4: a) Camera image of the sample and cantilever before acquisition of topography images and force vs distance curves on the TT shell embedded in resin. b) Topography image of TT surface. c) Young's modulus map of the area located by the blue square of panel b). The values were calculated by analyzing the force curves obtained by using the AFM in the force-volume mode. d) Distribution of the obtained Young's modulus values reported in the map shown in c).*

Each force-displacement curve was analyzed by implementing the Hertz model in MATLAB and using the approach curve for the fitting procedure. The corresponding Young's modulus value was plotted in the location where the curve was acquired, as shown by the color map of Figure 4 c). The moduli were calculated from the reduced ones by using the properties of the diamond tip ($E = 1140$ GPa and $\nu = 0.2$) and by assuming the Poisson's ratio of the shell approximately equal to $\nu = 0.3$, obtaining a rather large dispersion of the values. This is due to the underlying microstructure of the shell, as clearly highlighted by the SEM measurements. Panel d) reports the distribution of the obtained values, whose average is in the expected range (49) and equal to $E = 43 \pm 17$ GPa. No significant variations, within the error bars, are observed along the cross section of the shell. The underlying microstructure, and the consequent dispersion of the obtained values, hinder the possibility of detecting, on such length scales, local variations of the mechanical properties of the shell in the different layers because the AFM is probing the $CaCO_3$ crystals along with the lamellar interfaces. On the other hand, however, we know that it is the composite (9) and multiscale structure (20), according to which the lamellae are placed with different relative orientations, the reason of their optimized impact-resistance properties. Therefore, the macroscopic mechanical properties of the shell are the results of a complex, multiscale microstructure, that is not just the sum of the single components and that gives rise here to a wide dispersion of data in the force-displacement curve results.

Topography scans on smaller areas, reported in the Supplemental material, were able to reveal the underlying microstructure. A Young's modulus map on a $2 \times 2\ \mu m^2$ area of a *T. tricarinata* sample TC1, shows that there is no clear spatial correlation between the topographic features and the Young's modulus values (Figure S3). However, the effect of the microstructure is reflected also in this case in the rather large dispersion of data. In this case, the obtained value was $E = 48 \pm 17$ GPa. The standard deviation of the distribution is the same as that obtained on the $36 \times 36\ \mu m^2$ area shown in Figure 4. Since the dispersion of data is considerable, the difference with respect to the previous sample is not statistically significant. Therefore, the AFM investigation is not essential to determine the effective Young's modulus of the structure, but rather to highlight its inhomogeneity at this scale due to the lamellar structure (revealed by the SEM measurements) and to have a first guess for determining the mechanical properties of the overall system.

Topography scans with a probe specifically designed for imaging (Bruker RTESPA-300, nominal curvature radius = 8 nm) were also performed to gain a clearer idea of the appearance of the underlying microstructure, even in the case of very smooth surfaces. An example is reported in Figure S4 (Supplemental material). The topography, but especially the phase, show very well that the material is formed by many elements that, as seen in the SEM images, have a lamellar form but that here appear as sort of "tiles" or grains forming the structure.

### 3.2. Nanoindentation results

To compare the previous results with those obtained by a different characterization technique at the nanoscale, instrumented nanoindentation experiments were also conducted. Here, the main difference is that the force is not applied by means of a cantilever, but vertically, thus providing very accurate measurements when probing bulk or porous regions, like the interface between different lamellae. Although it is not possible to obtain the topography of the surface with this method, nanoindentation allows to obtain an estimation of the mechanical properties of the single elements forming the microstructure, while the AFM, as reported above, rather highlights the complex microstructure.

No significant gradients of the mechanical properties were observed in the analyzed cross sections of the shells (Figure 5). As shown below, this supports the approximation used in FEA simulations of considering constant homogeneous mechanical properties over the whole sample. In the indentation region, empty spaces were found that indicate a similar "porous" structure (Figure 5 and Figure S6), due to the microstructure of the shell. The analysis was then focused on the edge between the resin and the shell cross section.

As reported in Figure 6, the overall values of the Young's modulus and hardness were not significatively different among TT0, TT, and TC1 samples, which all displayed a rather large scatter in the data due to the porosity and roughness of the surface. In this case, it is possible that such dispersion screens small differences in the values of the mechanical properties among the different samples, thus requiring a more precise technique.

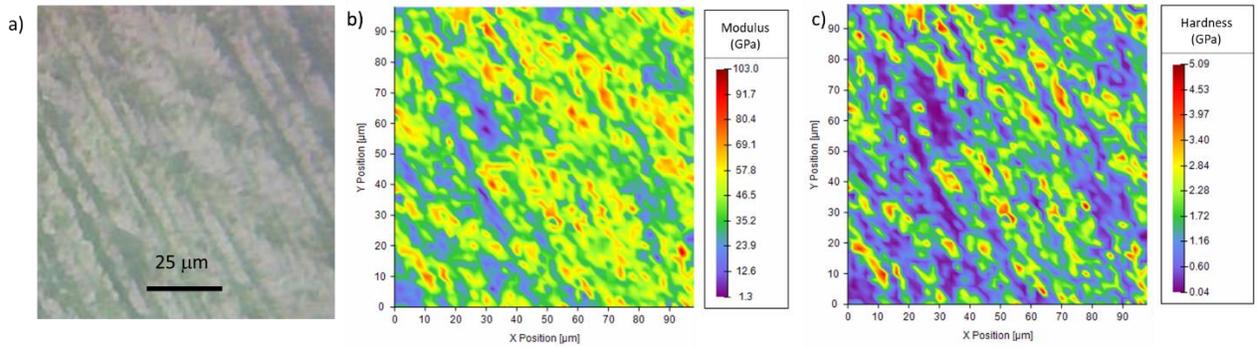

*Figure 5: a) Representative light microscopy image of the region of interest for which are reported b) Young's modulus and c) Hardness map. The sample depicted here is TT0. For these experiments the maximal load of 0.1 mN was used.*

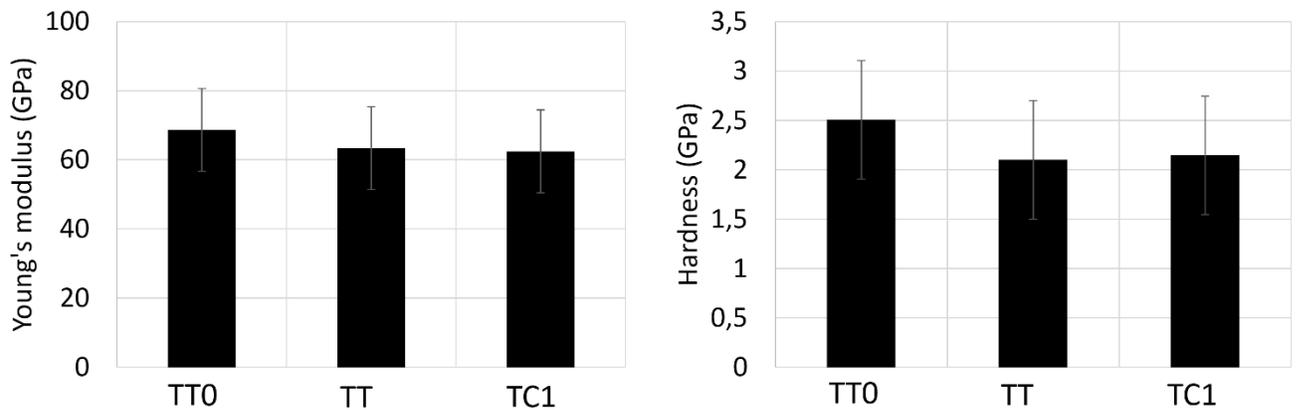

*Figure 6: Young's modulus and Hardness of the tested TT and TC samples. These values were selected from the maps reported in Figure S6 considering values with Hardness > 1.5 GPa, and Young's modulus > 50 GPa to avoid the empty regions. No significative differences emerge among these values.*

### 3.3. Ultrasound experiments results

Panels a)-d) of Figure 7: report, for some of the main observed resonance peaks, the normalized experimental modal shape along the longitudinal axis of the shell. This was obtained by plotting, at the frequency of the selected resonance mode, the imaginary part of the Fast Fourier Transform (FFT) as a function of the laser spot position, reported in the vertical axis as mm, that spanned over the entire length of the shell. This allowed to appreciate the amount of bending of the structure as a function of the position. The maximum value was normalized to one. Since out of plane vibrations were probed, the reported modes are predominantly bending modes.

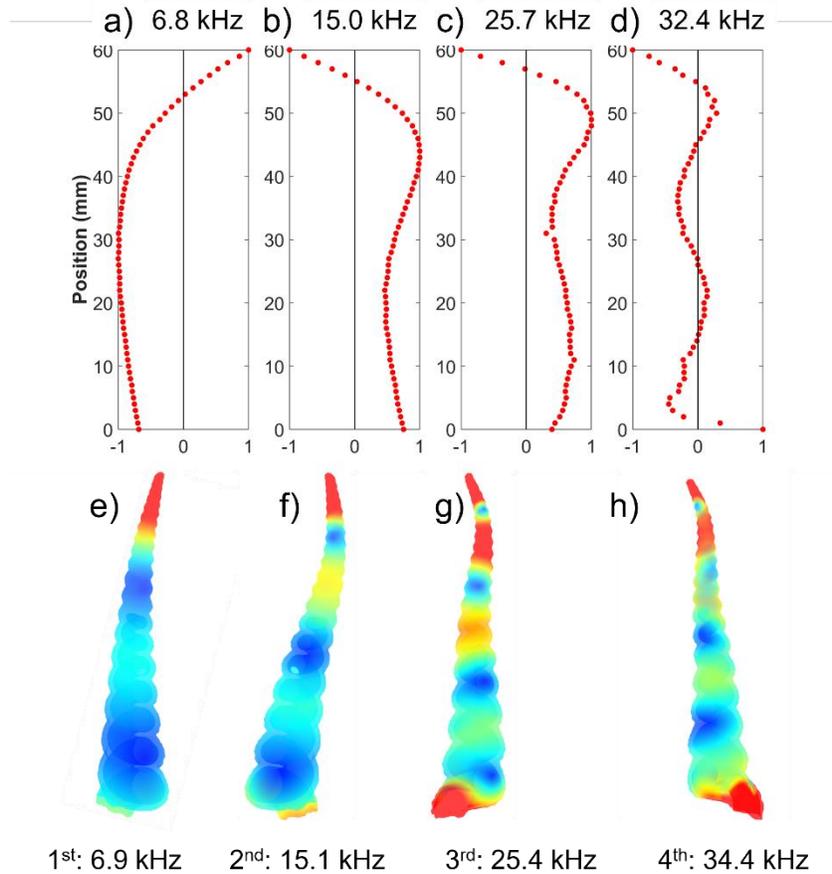

*Figure 7: a)-d) experimental normalized modal shapes for some of the main resonance frequencies for the TT. e)-h) vibration modes obtained by finite element simulations after optimization of the material parameters (see main text for details).*

Panels e)-h) of Figure 7 show the corresponding eigenmodes obtained by FEA using the material parameters that best match the experiments, as explained below in more detail. Considering that, from the experimental point of view, only bending modes can be detected, there is a good agreement between experiments and simulations both in terms of the modal shapes and corresponding resonant frequencies.

Similar measurements were also performed on two samples of *Turritellinella tricarinata*, TC1 and TC2, in the 5-100 kHz range. Results are reported in Figures S7 and S8. In these cases, some resonance peaks are visible at higher frequencies than observed in the previous sample, given the smaller size of these two shells.

## 4. Discussion

### 4.1. FEA-based Resonant Ultrasound Spectroscopy.

Resonance frequencies and mode shapes were calculated by using COMSOL Multiphysics 6.0 after importing the processed 3D geometry obtained from the CT scan. The eigenfrequencies were computed as a function of the Young's modulus and Poisson's ratio of the material by using the Structural Mechanics module and by performing an eigenfrequency study with parametric sweeps. The range of the swept parameters was initially chosen to include, as first guess, the values obtained from the mechanical characterization at the nanoscale. The calculated eigenfrequencies and relevant mode shapes were then compared with the experimental ones (Figure 8 a. A certain number $n$ of eigenmodes was considered and the optimized material parameters were determined as those minimizing the function:

$$Z = \log\left(\sum_{i=1}^{n}\left|\frac{\left(f_i^{exp} - f_i^{sim}\right)}{f_i^{exp}}\right|\right), \tag{1}$$

where $f_i^{exp}$ and $f_i^{sim}$ are the experimental and calculated eigenfrequencies of the *i*-th mode, respectively. The abscissae and ordinates of the color map reported in Figure 8b show, for the TT sample, the investigated range of Young's modulus and Poisson's ratio values, respectively, while the colors indicate the value of $Z$, which in this case has been computed by considering the first four resonance modes ($n = 4$). The red dot locates the minimum $Z$ value corresponding to the optimized parameters that, in this case, turn out to be about $E = 53 \pm 2$ GPa and $\nu = 0.35 \pm 0.06$, where the uncertainties are calculated by propagating the error of $f_i^{exp}$ and $f_i^{sim}$ to $Z$, respectively. Figure 8c compares the experimental and simulated eigenfrequencies. The broken symbols indicate simulated modes with a dominating longitudinal displacement which, therefore, cannot be efficiently experimentally detected by the vibrometer, sensitive to the out-of-plane displacements only. Notice that there are two such modes, as the second, around 33 kHz, is almost superimposed to another one.

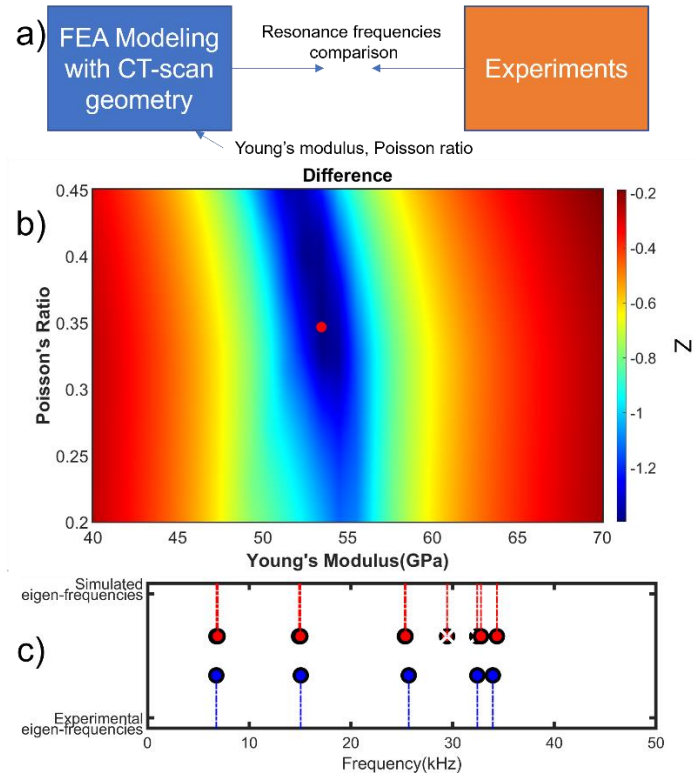

*Figure 8: a): Schematic of the comparison between FEA and experiments. b): color map of the difference between experiment and simulation for the first four resonance modes as a function of the material parameters (Young's modulus and Poisson ratio) for the TT shell. The red dot represents the point that*

*matches experiments and simulation for all the resonances considered. c): comparison between experimental and simulated eigenfrequencies.*

In the case of the two *T. tricarinata* samples, three frequencies were considered in the minimization process, from which we obtained $E = 70 \pm 2$ GPa and $\nu = 0.32 \pm 0.05$ for TC1 and $E = 61 \pm 3$ GPa and $\nu = 0.31 \pm 0.09$ for TC2, respectively. Results are reported in the Figures S9 and S10. By looking at images shown in Figure 8 b), S9 a) and S10 a), we can notice that the variation of $Z$ is greater when driven by the Young's modulus rather than by the Poisson ratio. Thus, the Young's modulus is a much more relevant parameter in the inversion procedure for determining the mechanical properties of the samples, indicating that only this quantity should be compared at the different investigated length scales. The Young's modulus values obtained in the three cases are all within the range determined by the nanoindentation experiments (Figure 6), but we can now appreciate some differences, as dissimilarities in the mechanical properties, that can be due to individual variability. Indeed, it is known that environmental and ecological conditions, like sea temperature and oxygen concentration, may affect the elemental composition and isotopic ratio in the mollusk shell (50–53). This could be the rationale for a future work, in which a large amount of specimens of the same species but with different ecological history could be analyzed.

### 4.2. Frequency domain analysis

Finally, to better understand the dynamic behavior of the structure of the *Turritella* shells, we performed a comparative numerical frequency domain analysis using the obtained mechanical data. The actual geometry, obtained from the CT scan, was used to compute the frequency response spectrum in a certain frequency range. The damping was modelled using an isotropic loss factor $\eta_s$ (defined as the ratio of energy dissipated in unit volume per radian of oscillation to the maximum strain energy per unit volume). This value was estimated by comparing the simulated particle velocity field as a function of frequency with the experimentally-derived one in the range (0.5-100) kHz. The computed spectra are reported in Figure 9 for different loss factor values, along with the experimental data for sample TC1. It is apparent that the addition of damping considerably improves the agreement with the experimental data, although there is some discrepancy between peak amplitudes at some resonance frequencies. For example, the experimental peak at 20 kHz is rather damped, but the one at about 65 kHz is still clearly noticeable. This can be related to the out of plane nature of the measurements, which can introduce some differences in the measured relative peak intensities, as well as in the number of detected peaks. A loss factor of $\eta_s \approx 5 \times 10^{-3} - 1 \times 10^{-2}$ reproduces the experimental data well, since for larger values (as in the case of $\eta_s = 3 \times 10^{-2}$, also shown in the figure) some spectral features are lost.

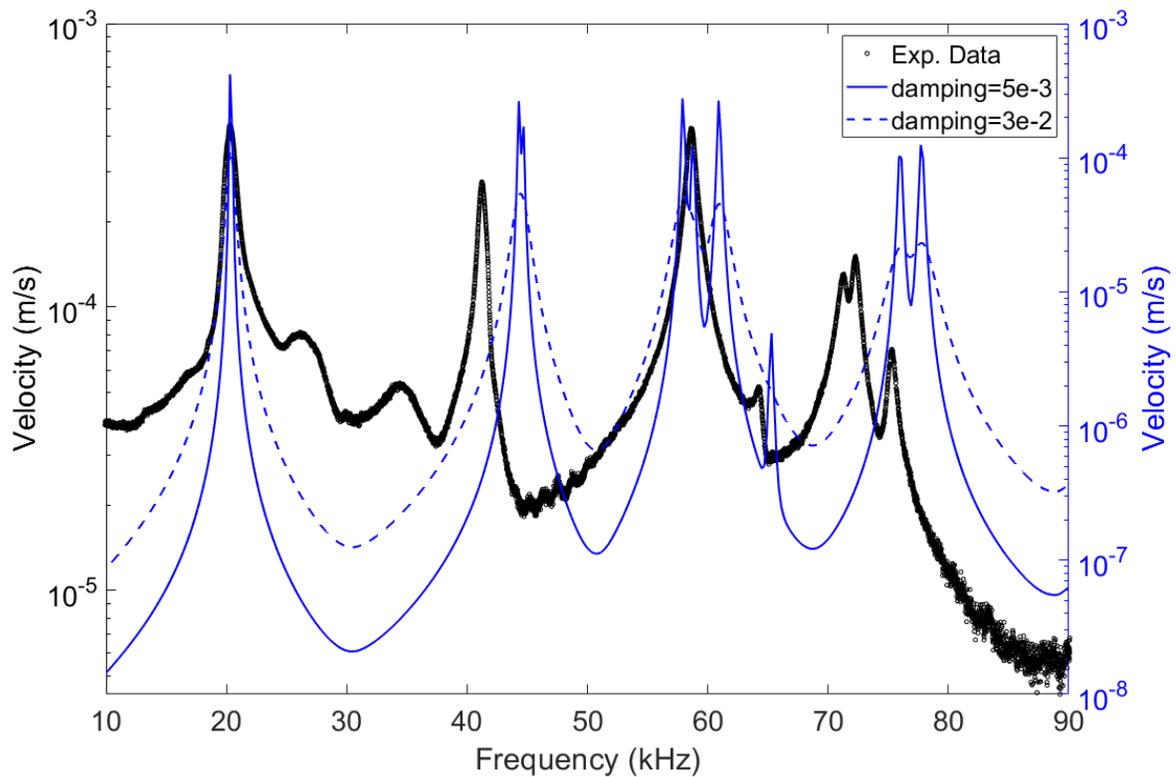

*Figure 9: Numerical frequency domain analysis. Black symbols, left axis: Experimental frequency response spectrum of the* Turritellinella *sample TC1. Solid and dashed blue lines, right axis: numerical spectra for loss factors equal to $\eta_s = 5 \times 10^{-3}$ and $\eta_s = 3 \times 10^{-2}$, respectively.*

To understand the influence of the full *Turritella* structure on its modal vibration properties, we then compared its frequency response with that of a similar, but simpler, ideal structure. The *Turritella* structure is helicoconic and the internal hollow spiral structure, in comparison with a basic conical one, can presumably provide better structural strength and impact damping to protect the living organism from predators. Moreover, it also presents a hierarchical organization, since every spiral contains, on its surface, several thinner spirals that curl around the larger ones. The main reason for this feature, similar to a drill, is most probably to help the organism to burrow under the mud (54). Therefore, it is interesting to compare this structure to its simplest analog, where only the conical shape has been retained. This can help to better understand the influence on the mechanical behavior of "second order" elements in the hierarchical structure. The geometry of the conical structure was designed to have the same mass and length as the *Turritella* shell used for the comparison. The density, Young's modulus, Poisson ratio, and loss factor were the same in both cases. Figure 10 reports the vibration strain energy as a function of the frequency for these two systems. Results show that, for the two structures, the modes at frequencies up to 80 kHz are essentially the same. This is clearly due to the common basic structural features, i.e., the conical shape. Panels b), c) and e), f) report the first two simulated modal shapes for the cone and the *Turritella*, respectively. Arrows indicating the displacement magnitude and direction are also reported to help the comparison. One can easily see that these two modes present the same vibration characteristics.

On the other hand, at higher frequencies, the greater structural complexity of the *Turritella* gives rise to a richer behavior, with more numerous peaks. These are related to the spiral, hierarchical structure of the real shell, with several more modes occurring above 100 kHz. Moreover, by comparing modes that are rather close in frequency, we can observe a more complex deformation behavior for the shell. For example, panels d) and g) show two high-frequency modes for the cone and the shell at about 162 kHz and 166 kHz, respectively. Both modes reported in panel d) and g) feature a mixed behavior but, while the former, belonging to the cone, has a dominant bending contribution along with torsional components, the mode of the *Turritella* also displays non-negligible longitudinal components, indicated by the vertical arrows, together with more pronounced torsional effects than for the conical structure.

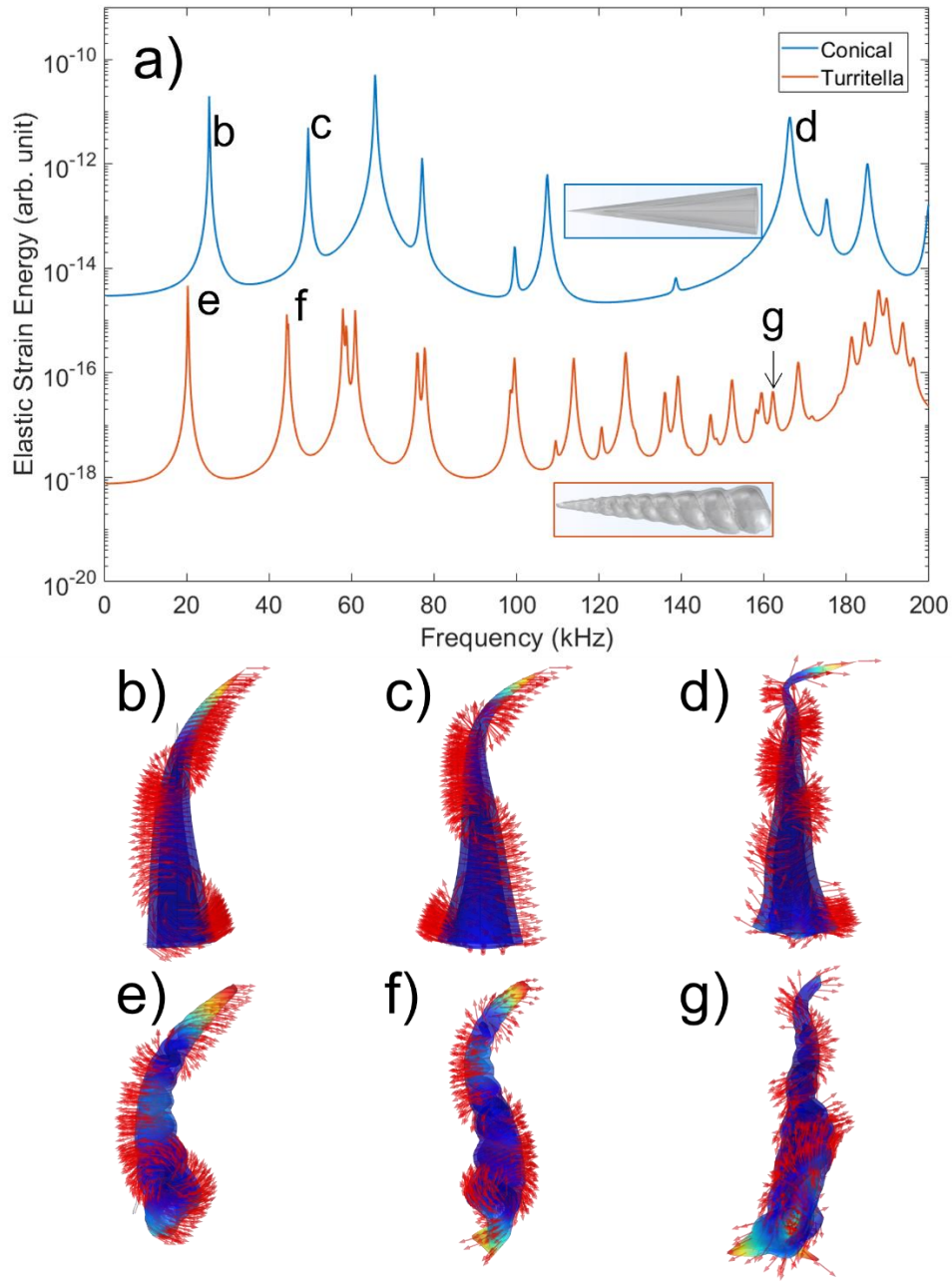

*Figure 10: a) Calculated elastic strain energy spectrum for the* Turritellinella *sample TC1 (red) and for an ideal, conical shell (blue) with the same mass and length as the* Turritella*. The loss factor, in both cases, is equal to* $5 \times 10^{-3}$*. b)-d) Numerical modal shapes for the conical shell geometry. Arrows represent displacement. b) First mode at 25.5 kHz, c) Second mode at 49.5 kHz, d) Higher frequency mode at 166.1 kHz. e)-g) Numerical modal shapes for the* Turritella *shell geometry. Arrows represent displacement. e) First mode at 20.3 kHz, f) Second mode at 44.3 kHz, g) Higher frequency mode at 161.9 kHz.*

In general, we can speculate that this behavior can help to achieve a better performance in terms of impact absorption, since a higher modal density i.e., a larger number of modes in the same frequency range, featured by the *Turritella* shell structure, indicates the presence of more equivalent spring/damper subsystems in the system and thus a greater efficiency in the capability of storing (and dissipating) strain energy, and therefore possibly a more resilient structure to impacts. It is interesting to observe that organisms with orthoconical shells belonging to the same phylum existed in the past (for example, the extinct nautiloid cephalopod *Orthoceras* (55)), but these were part of a taxon of nektonic animals, occupying a different biotype from that of the *Turritella*.

5. Conclusions

In this paper, we have proposed a procedure that integrates microstructural analysis and FEA-based RUS to investigate the mechanical and vibrational properties of biological materials, implementing it on two varieties of seashells. The method allows to perform RUS on biological samples with complex shape and to obtain information not only regarding the frequency spectrum of the various vibrational modes, but also on their specific modal shapes, thus rendering a comparison with numerical data unambiguous. Characterization at the nanoscale reveals the extraordinarily complex microstructure of the shells. In general, this type of analysis can be useful to check the presence of heterogeneity, grading, anisotropy, and other mechanical features that might need to be included in macroscopic simulations, such as anisotropic moduli, multi-material components or with spatially graded properties. In this study, it was possible to use homogeneous mechanical properties for the entire structure, but in general, to fill the gap between the properties at the microscale and those at the macroscale, multiscale structural models should be designed and investigated, as has been done for other natural systems (56). Moreover, homogenization procedures might also be probably necessary. On the other hand, the results provide a starting guess for the macroscopic parameters to be introduced in the simulations. Here, we did not observe a significant change of the properties along the cross section of the samples, but the large dispersion of data in the AFM results is an intrinsic effect of the shell microstructure which turned out to be the dominant effect. From the nanoindentation experiments we could obtain the correct order of magnitude of the Young's modulus to insert as a starting value for the FEA. The FEA simulations were performed on the actual shell geometry, as obtained from the CT scan. From the comparison with the experimental resonance frequencies and modal shapes, the mechanical properties that describe the macroscopic behavior of the structure were determined. The obtained Young's modulus values are within the range obtained by the nanoindentation experiments, but differ slightly from sample to sample.

Finally, the dynamic response of the complex structure of the *Turritella* was compared to that of a simplified, purely conical, shell. It was shown that, in addition to the same vibration modes as the ideal conical one, the real shell features a much richer spectral behavior, deriving from the greater structural complexity. This could also (but clearly, not exclusively), be related to improved attenuation capabilities of the structure under impacts. It is known that many natural systems that are optimized for impact attenuation feature hierarchical, interfacial, porous and composite architectures (5). In the dynamical analysis reported here, we have shown that the hierarchical and helicoconic characteristics play a significant role in the rich frequency response of the shells. To verify this hypothesis, the method presented here could be implemented in a systematic manner for the characterization of other biological structures of interest for their vibration behavior, focusing on hierarchical features, allowing to gain further insight into their evolutionary development, to reveal the effect of environmental conditions, like ecological stress, on different samples of shells of the same species and to draw inspiration for efficient impact resistant or vibration damping bioinspired designs.

## Acknowledgements

YL, ML, FSS, GG, VFDP, ASG, NMP, FB, MT acknowledge the European Commission under the FET Open ''Boheme'' Grant No. 863179. We thank the Science Museum of Trento (MUSE) for providing the *Turritella terebra* samples. Special thanks to Salvatore Guastella for the SEM images and to J. Pappas for useful discussions.

Supplemental material

SEM

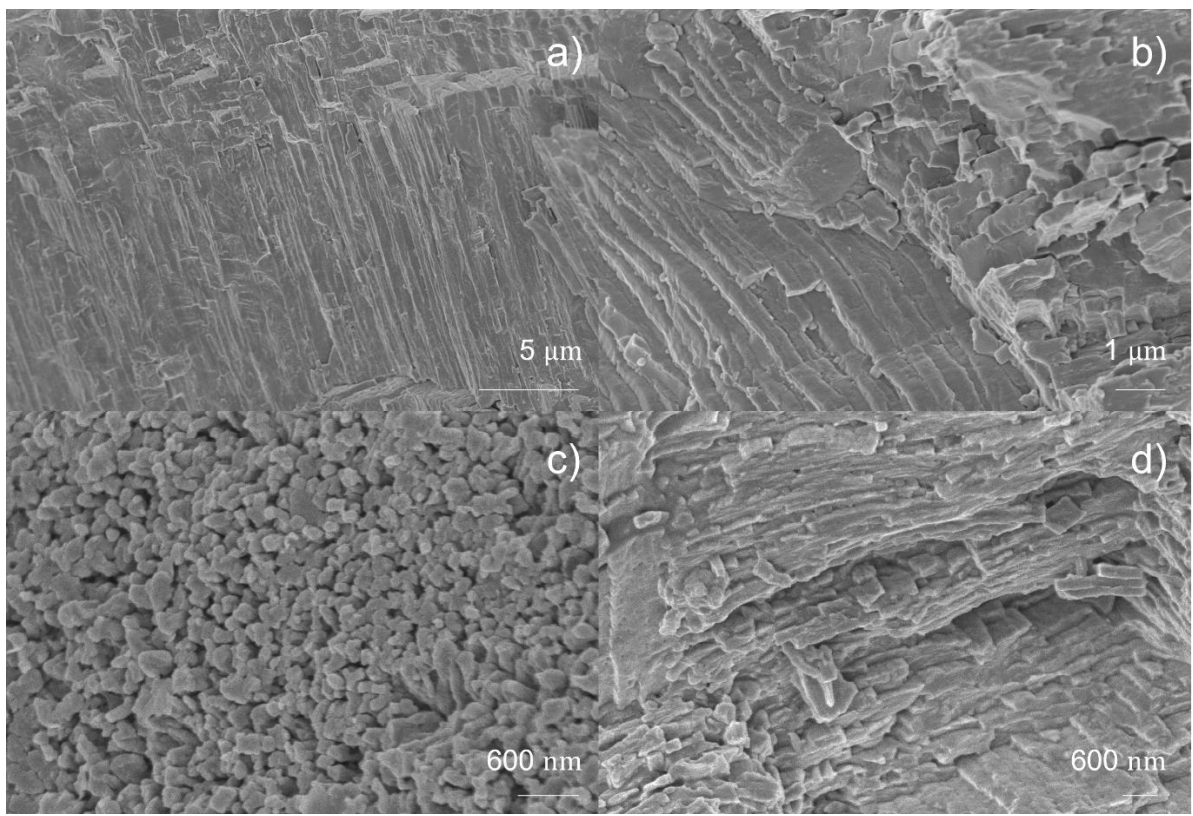

*Figure S1.SEM images of the* Turritellinella tricarinata *shell TC1. a)-b): second-order lamellae at different magnifications. c)-d): different views of the third-order lamellae.*

## Atomic Force Microscopy

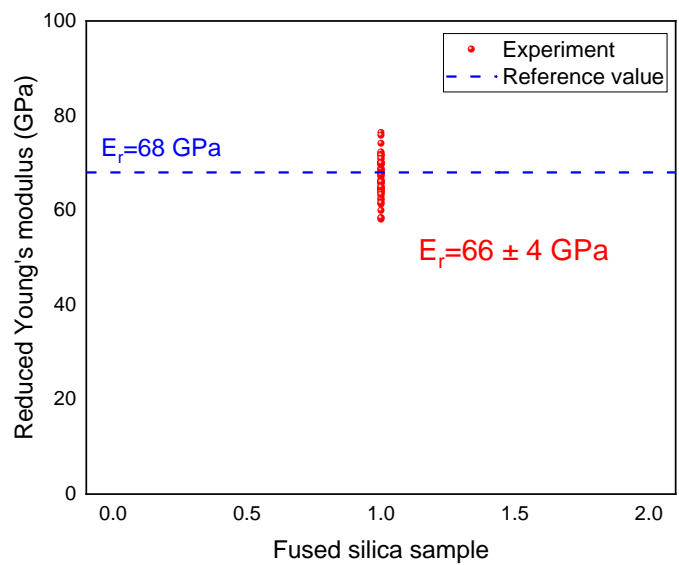

*Figure S2 Red symbols: experimental values of the reduced Young's modulus, $E_r$ obtained on a test fused silica sample by means of AFM force vs distance curves and fit with the Hertz model. The blue dashed line is the reference $E_r$ value for fused silica and diamond.*

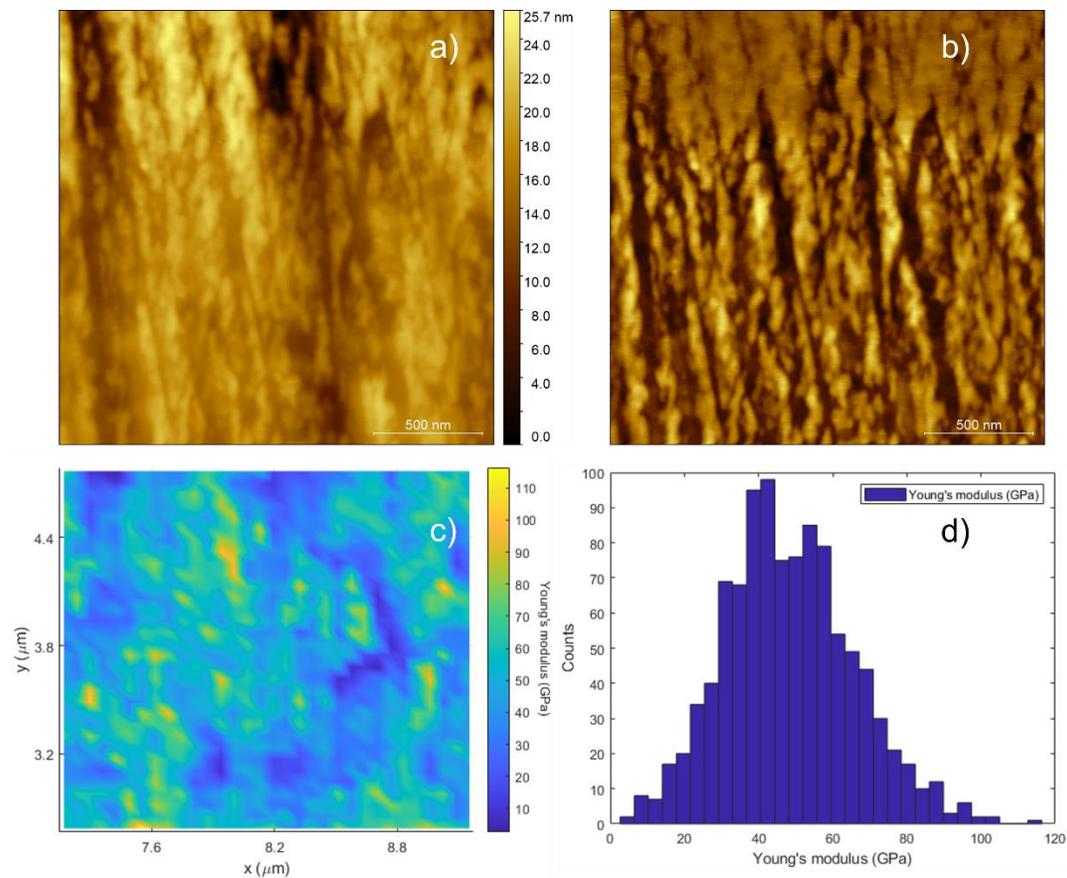

*Figure S3 Topography, a) and phase, b) images of a $2 \times 2\ \mu m^2$ polished surface of the* Turritellinella tricarinata *sample TC1. c) Young's modulus map of the area shown in a). d) Distribution of the obtained Young's modulus values reported in the map shown in c).*

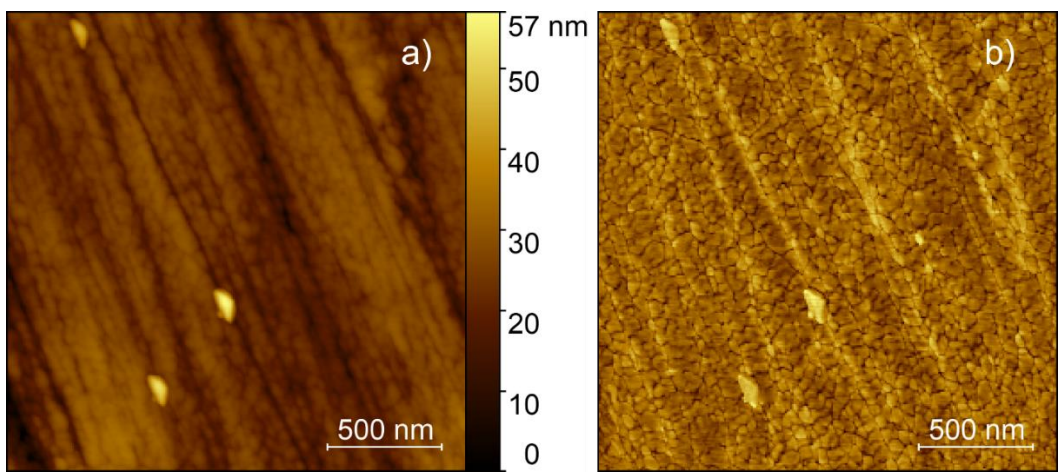

*Figure S4. Topography a) and phase b) image acquired, with an AFM imaging silicon probe on the polished surface of the inner layer of the* T. tricarinata *shell TC1.*

Instrumented nanoindentation

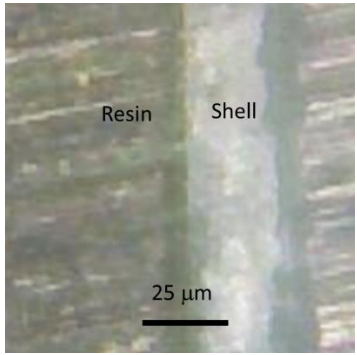

Figure S5: Resin shell cross section interface observed with light microscopy.

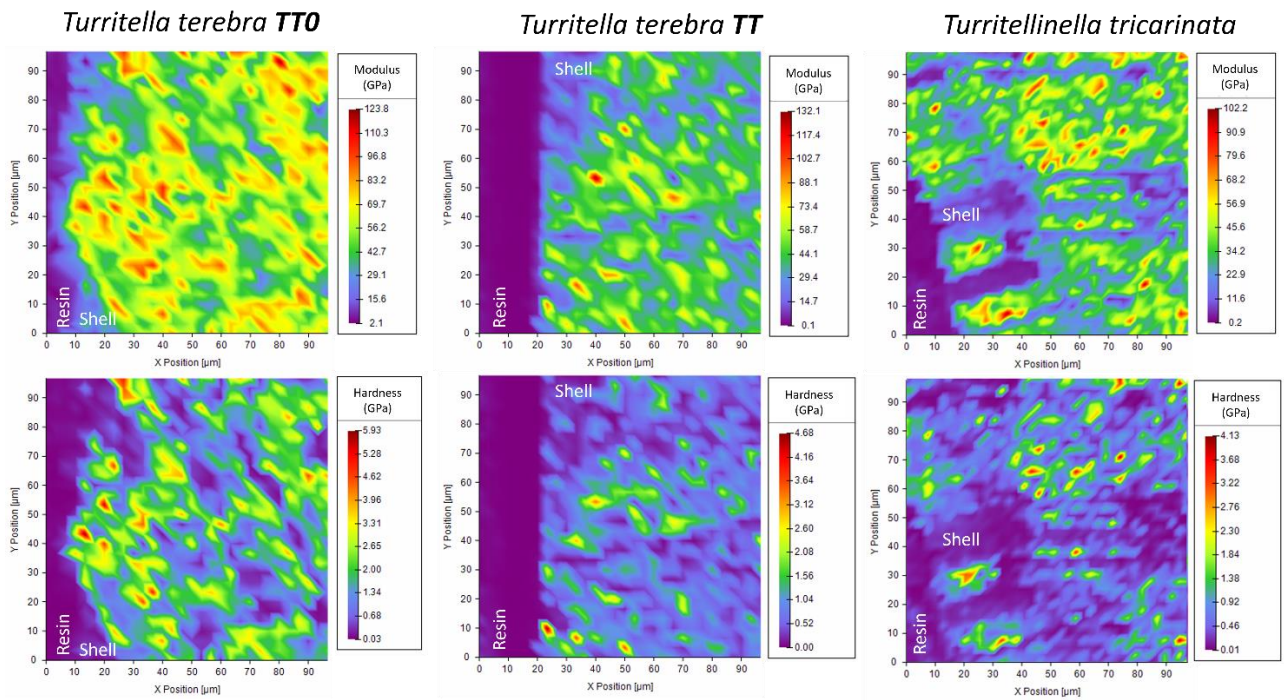

Figure S6: Representative maps of the mechanical properties of the 3 different types of samples analyzed. **Note that the scales are not the same.**

Ultrasonic measurements

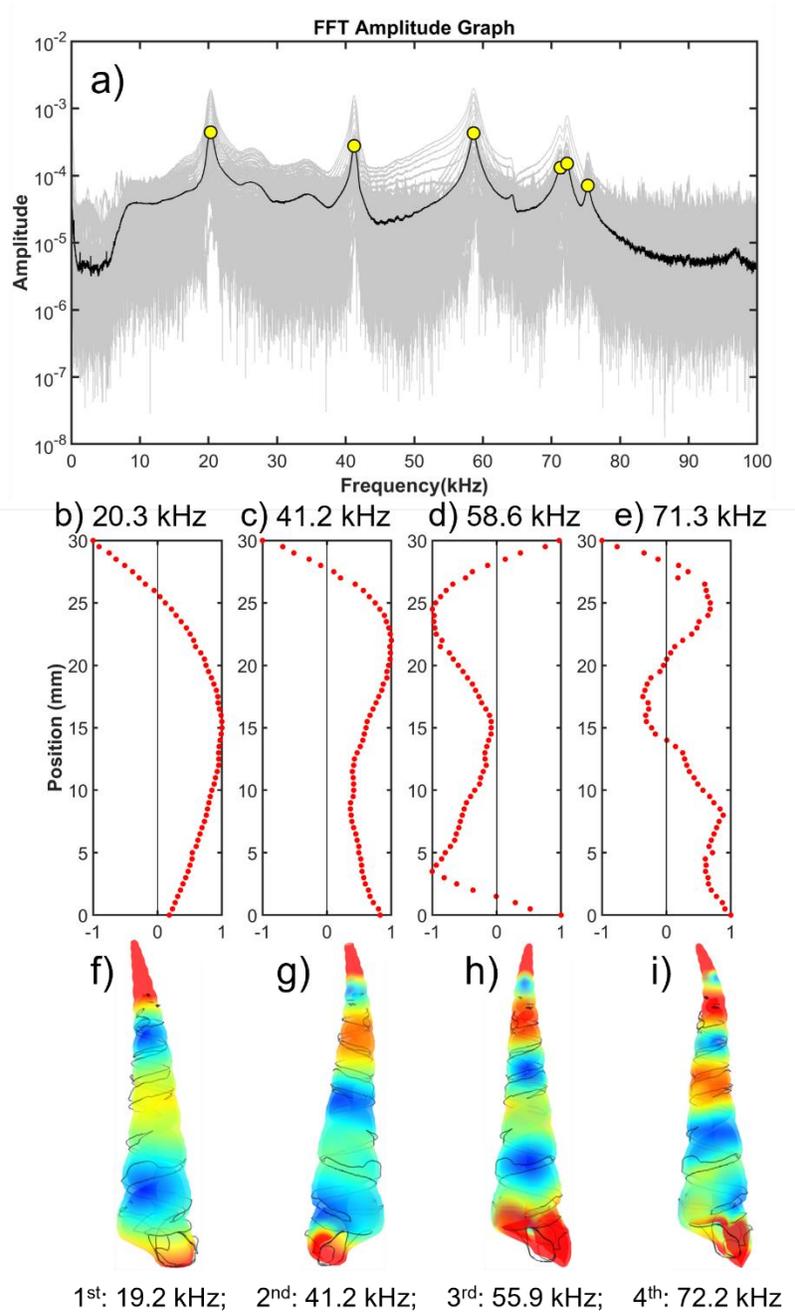

*Figure S7 Upper panel: gray lines are the frequency spectra recorded along the longitudinal axis of the* Turritellinella tricarinata *shell TC1 while the black line represents their average and the yellow symbols the position of the main peaks. Central panel: the normalized modal shapes for some of the main resonance frequencies for the shell. Lower panel, a)-d) shows the vibration modes obtained by finite elements simulations after optimization of the material parameters (see main text for details).*

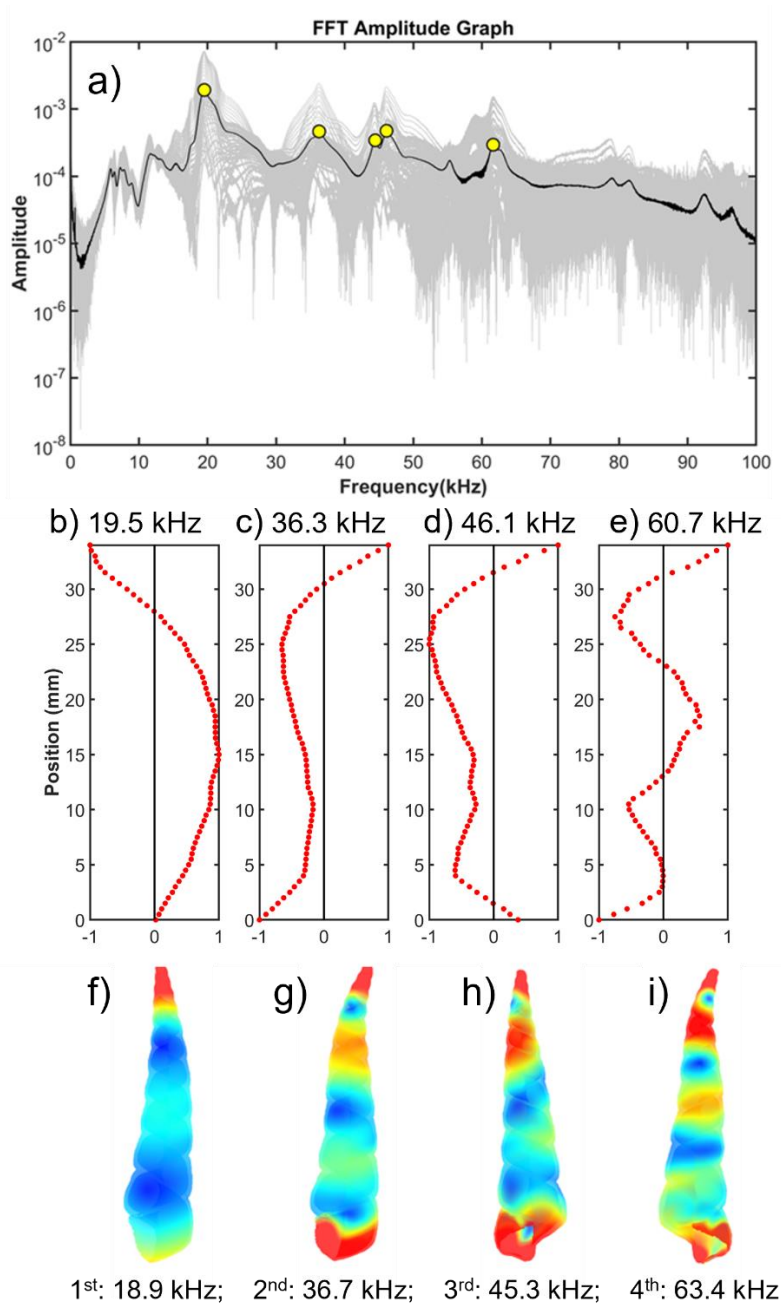

*Figure S8 Upper panel: gray lines are the frequency spectra recorded along the longitudinal axis of the* Turritellinella tricarinata *shell TC2 while the black line represents their average and the yellow dots the position of the main peaks. Central panel: the normalized modal shapes for some of the main resonance frequencies for the shell. Lower panel, a)-d) shows the vibration modes obtained by finite elements simulations after optimization of the material parameters (see main text for details).*

## FEA-based Resonant Ultrasound Spectroscopy

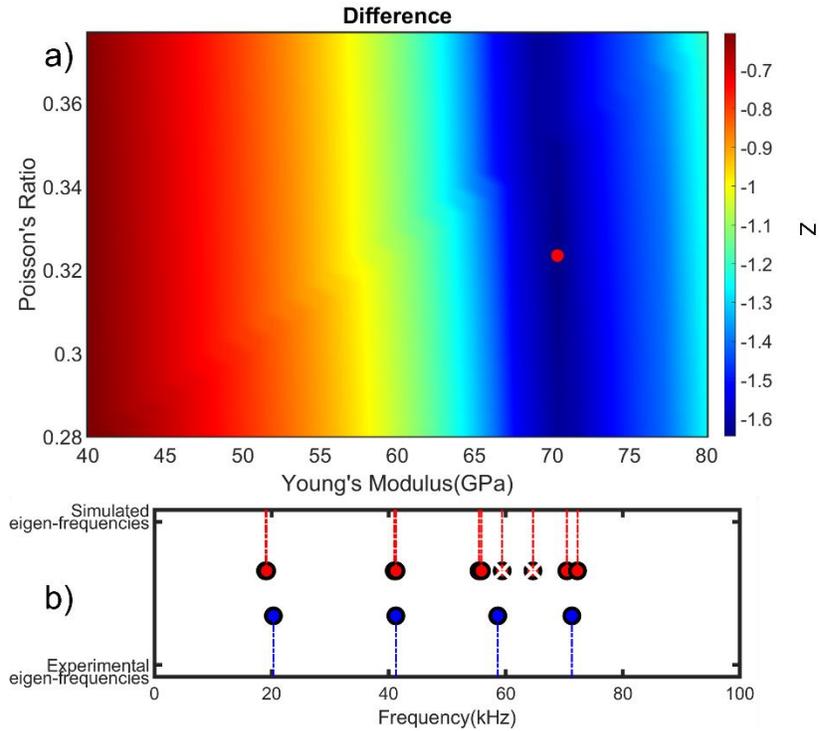

*Figure S9 Upper panel: color map of the difference between experiment and simulation for the first three resonance modes as a function of the material parameters (Young's modulus and Poisson ratio) for the* Turritellinella tricarinata *shell, sample TC1. The red dot represents the point that matches experiments and simulation for all the resonances considered. Lower panel: comparison between experimental and simulated eigenfrequencies.*

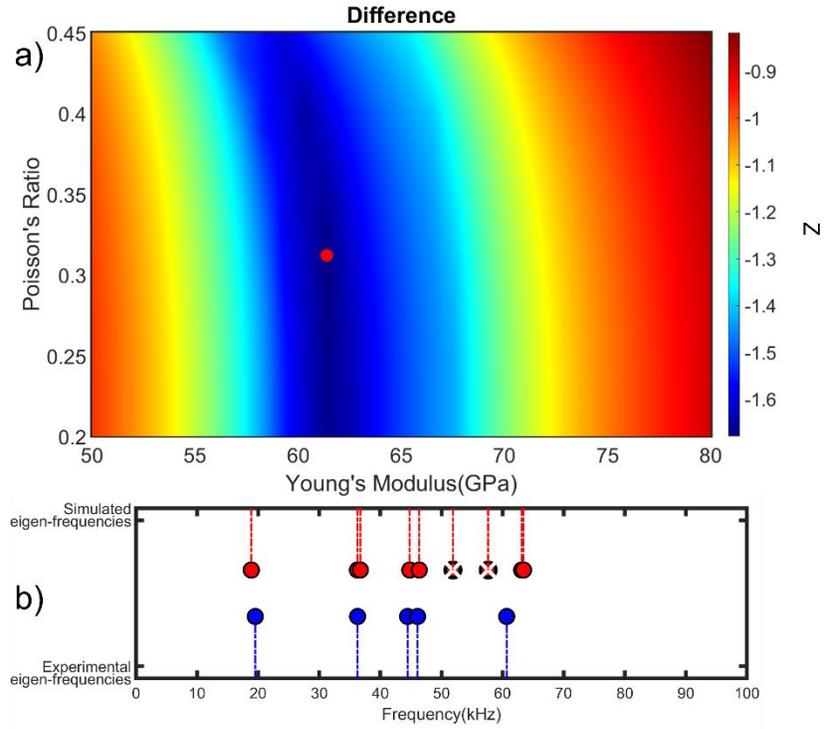

*Figure S10 Upper panel: color map of the difference between experiment and simulation for the first three resonance modes as a function of the material parameters (Young's modulus and Poisson ratio) for the* Turritellinella tricarinata *shell, sample TC2. The red dot represents the point that matches experiments and simulation for all the resonances considered. Lower panel: comparison between experimental and simulated eigenfrequencies.*